\documentclass[trackchanges]{aastex63}

\usepackage{multirow}
\usepackage{amsmath,amssymb}

\received{August 26, 2021}
\revised{February 3, 2022}
\accepted{March 5, 2022}
\submitjournal{AJ}

\turnoffedit

\begin{document}

\title{On the Effect of Stellar Activity on Low-resolution Transit Spectroscopy and the Use of High Resolution as Mitigation}

\correspondingauthor{Fr\'ed\'eric Genest}
\email{frederic.genest.1@umontreal.ca}

\author[0000-0003-0602-9106]{Fr\'ed\'eric Genest}
\affiliation{Institut de Recherche sur les Exoplan\`etes (iREx), \\Universit\'e de Montr\'eal, D\'epartement de physique, \\1375 Avenue Th\'er\`ese-Lavoie-Roux, Montr\'eal, QC, Canada}

\author[0000-0002-6780-4252]{David Lafreni\`ere}
\affiliation{Institut de Recherche sur les Exoplan\`etes (iREx), \\Universit\'e de Montr\'eal, D\'epartement de physique, \\1375 Avenue Th\'er\`ese-Lavoie-Roux, Montr\'eal, QC, Canada}

\author[0000-0001-9427-1642]{Anne Boucher}
\affiliation{Institut de Recherche sur les Exoplan\`etes (iREx), \\Universit\'e de Montr\'eal, D\'epartement de physique, \\1375 Avenue Th\'er\`ese-Lavoie-Roux, Montr\'eal, QC, Canada}

\author[0000-0002-7786-0661]{Antoine Darveau-Bernier}
\affiliation{Institut de Recherche sur les Exoplan\`etes (iREx), \\Universit\'e de Montr\'eal, D\'epartement de physique, \\1375 Avenue Th\'er\`ese-Lavoie-Roux, Montr\'eal, QC, Canada}

\author[0000-0001-5485-4675]{Ren\'e Doyon}
\affiliation{Institut de Recherche sur les Exoplan\`etes (iREx), \\Universit\'e de Montr\'eal, D\'epartement de physique, \\1375 Avenue Th\'er\`ese-Lavoie-Roux, Montr\'eal, QC, Canada}

\author[0000-0003-3506-5667]{\'Etienne Artigau}
\affiliation{Institut de Recherche sur les Exoplan\`etes (iREx), \\Universit\'e de Montr\'eal, D\'epartement de physique, \\1375 Avenue Th\'er\`ese-Lavoie-Roux, Montr\'eal, QC, Canada}

\author[0000-0003-4166-4121]{Neil Cook}
\affiliation{Institut de Recherche sur les Exoplan\`etes (iREx), \\Universit\'e de Montr\'eal, D\'epartement de physique, \\1375 Avenue Th\'er\`ese-Lavoie-Roux, Montr\'eal, QC, Canada}

\begin{abstract}

We present models designed to quantify the effects of stellar activity on exoplanet transit spectroscopy and atmospheric characterization at low ($R=100$) and high ($R=100,000$) spectral resolution. 
We study three model classes mirroring planetary system archetypes: a hot Jupiter around an early-K star (HD 189733 b); a mini-Neptune around an early-M dwarf (K2-18 b); and terrestrial planets around a late M dwarf (TRAPPIST-1). We map photospheres with temperatures and radial velocities (RV) and integrate specific intensity stellar models. We obtain transit spectra affected by stellar contamination, the Rossiter--McLaughlin effect (RME), and center-to-limb variations (CLV). We find that, at low resolution, for later-type stars, planetary water features become difficult to distinguish from contamination. Many distributions of unocculted active regions can induce planetary-like features of similar amplitudes in the case of a late M dwarf. 
Atmospheric characterization of planets around late-type stars will likely continue to suffer from degeneracy with stellar activity unless active regions' parameters can be constrained using additional information.
For the early-K star, stellar contamination mostly manifests itself through a slope at optical wavelengths similar to Rayleigh scattering. 
In all cases, contamination induces offsets in measured planet radii. 
At high resolution, we show that we can determine the origin of $\text{H}_2$O and CO detection signals and lift the degeneracy observed at low resolution, provided sufficient planet RV variation during transit and adequate correction for the RME and CLV when required.
High-resolution spectroscopy may therefore help resolve issues arising from stellar contamination for favorable systems.

\end{abstract}



\section{Introduction}
\label{sec:introduction}

Since the first detection of HD 209458 b's transit two decades ago \citep{Charbonneau2000}, exoplanet transits have become one of the most common ways to detect and characterize exoplanets. There is now a wealth of known transiting planets ranging from terrestrial planets to hot Jupiters available for atmospheric characterization via transit and eclipse spectroscopy, most of which were discovered thanks to the Kepler satellite \citep{Borucki2010}, while many more are expected from the ongoing TESS mission \citep{Ricker2015}.

Atoms such as Na, K, Fe, and H have been detected in the atmosphere of many extrasolar planets using transit observations in the optical. Detections have been made both at low resolution from space \citep[for example with HST/STIS;][]{Charbonneau2002} and high resolution from the ground \citep[with HARPS and ESPRESSO among others;][]{Casasayas2019, Chen2020, Gibson2020}. Starting more recently, molecules such as $\text{H}_2$O and CO have been detected in the near infrared transmission spectra of a number of exoplanets, again both at low (in most cases thanks to HST/WFC3; \citet{Deming2013, McCullough2014, Benneke2019}; see also \citet{Iyer2016, Sing2016} for reviews of many data sets) and high spectral resolution \citep[with instruments such as CRIRES and CARMENES;][]{Snellen2010, Brogi2016, Alonso-Floriano2019}. High-resolution detections are made possible by exploiting the change in radial velocity of the planet during transit, as long as the amplitude of that change is large enough to be detected by the instrument used.

However, not all observed planets yield atmospheric detections, even when a significant signal is expected. GJ 1214~b \citep{Berta2012} is a well-known example. Even in cases with detections, it is common for the observed features to be weaker than expected from atmosphere models \citep{Sing2011, Deming2013}. Common explanations involve the presence of clouds or hazes high in the atmosphere \citep{Iyer2016, Sing2016}: the atmosphere would be totally or partially opaque below a certain altitude, mimicking the appearance of a planet with far fewer absorbers than there actually are.

Unfortunately, clouds and hazes are not the only possible source of confusion in transmission spectra. Since over a decade ago, stellar activity has been recognized as a likely source of contamination in transit observations \citep{Pont2007, Pont2013, Oshagh2014}. Inhomogeneities on the stellar surface, both cold spots and bright plages (faculae), can introduce differences in the spectrum emerging from the transit chord compared to the spectrum from the entire surface. \citeauthor{Rackham2018} call this phenomenon the \textit{transit light source effect}. This can introduce systematic errors in the measurements of the wavelength dependent planet radius $R_p (\lambda)$ such as an offset, as well as\edit2{ atomic and} molecular absorption and emission features in the transmission spectrum, depending on the coverage fraction, distribution, and relative spectra of inhomogeneities. For example, unocculted dark spots on a late type star are expected to induce water absorption features in a planet's transmission spectrum.

Extensive efforts have recently been undertaken to model the effects of stellar contamination on transmission spectra for a wide range of host star spectral types (F to M) and spot/faculae covering fractions \citep{Rackham2018, Rackham2019}. \citeauthor{Rackham2018} have computed the resulting photometric variation for all their models and the wavelength-dependent modulation of the observed transmission spectra at low spectral resolution. The expected magnitude of those effects varies greatly across the range of host star spectral types according to these models, generally becoming more significant for later spectral types.

Compared to how much is known about the Sun's activity and surface inhomogeneities, knowledge about the activity of general stars, especially late-type stars like M dwarfs, is limited. Many stars are expected to be active based on magnetic activity, flares, XUV and X-ray fluxes, and other indicators, but the coverage fractions of spots and faculae are still not constrained very well for most stars. Photometric monitoring can provide estimates of these coverage fractions, but the uncertainties are large and the observed variability is degenerate with the set of possible inhomogeneity distributions. In some cases, spot crossing events can be detected during transit and used to infer some properties of said spots, like spot size and temperature contrast \citep{Sing2011, Espinoza2019}. Doppler imaging of active stars has been conducted on multiple occasions \citep{Barnes2015, Barnes2017}, but the resolution of such techniques is not sufficient to put tight constraints on spot sizes, temperatures, covering fractions, etc. We point to \citet{Strassmeier2009} for a review on the topic of starspots.

Solving the problem of stellar contamination at lower spectral resolutions can be very challenging or almost impossible because of the amount of degeneracy between spot distributions, temperatures, atmosphere composition, clouds, etc. 
In this work, we turn to the very high spectral resolutions achievable with many modern optical and near infrared spectrographs (e.g. $R \gtrsim 50,000$) and investigate if they can provide a solution to the problem.
The planetary atmosphere and stellar inhomogeneities present different radial velocity variations during a transit. Therefore, in principle, these two signals should be distinguishable in a cross-correlation analysis. One way to conduct such an analysis is to compute correlation maps of the transmission spectra times series with models, as a function of assumed systemic velocities and planetary orbital velocities. In the work below, we will thus model the effects of stellar activity, rotation, and limb darkening on high-resolution transit spectra and verify if planetary features can be safely disentangled from stellar systematic effects at such resolutions.

As other efforts have pointed out before \citep{Brogi2016, Casasayas2019, Casasayas2020, Chen2020}, in many cases, stellar rotation must be carefully accounted for when analyzing high-resolution transit spectra. This is mainly because of the Rossiter--McLaughlin effect \citep[RME;][]{McLaughlin1924, Rossiter1924, Triaud2018}: stellar lines profiles can be significantly affected during transit, in turn leaving an imprint on a cross-correlation analysis conducted to look for a planetary signal. 
This becomes a concern when lines from the planetary atmosphere also appear in the stellar spectrum. 
Convective blueshift and its inhibition in active regions is another phenomenon related to stellar surface radial velocities that can affect line profiles during transit \citep{Dumusque2014}. 
Additionally, the center-to-limb variation (CLV) of the emergent stellar spectrum can vary as a function of wavelength in spectral lines \citep{Czesla2015}, which can also affect the observed planetary line profiles.
All of these issues will have to be included in our analysis.

This work will focus on three archetypal systems: a hot Jupiter such as HD 189733 b orbiting an early-K dwarf \citep{Bouchy2015}, a super-Earth/mini-Neptune orbiting an early-M dwarf like the K2-18 system \citep{Montet2015, Benneke2017} inspired by the recent HST/WFC3 water detection on K2-18 b \citep{Benneke2019}, and a terrestrial planet around a late active M dwarf as was discovered in the TRAPPIST-1 system \citep{Gillon2017}.

The structure of the code, model parameters and analysis methods will be presented in section \ref{sec:Methods}. Our main results can be found in section \ref{sec:Results}. We discuss the possible implications of these results in section \ref{sec:Discussion}. Section \ref{sec:Conclusion} offers a summary of our findings and improvements to be made in future works.


\section{Methods}
\label{sec:Methods}

Our first goal in this study is to model transit spectra as accurately as possible while taking into account the effects of stellar rotation, CLV, convective blueshift, stellar activity, the distribution of active areas, and the orbital motion of the planet. The models are generated for three different cases of stellar contamination: (1) a planet with an atmosphere transiting a star without spots, to obtain pure atmospheric signatures; (2) a planet without an atmosphere transiting a star with surface activity, to isolate the effect of stellar contamination; and (3) a planet with an atmosphere and an active star, to combine both types of signatures,\edit2{ to} simulate possible observations, and try to separate the origin of the signals.

\subsection{Components of the Models}

The stellar spectra used to build the models are specific intensity spectra $I_{\lambda}(\mu)$ obtained with the PHOENIX code \citep[][Peter Hauschildt, private communication]{Hauschildt1999}. The use of specific intensity spectra ensures that CLV is treated adequately for each photospheric component. Three temperatures were assigned for each archetypal models: $T_{\text{phot}}$ for the quiet photosphere, $T_{\text{spot}}$ for dark spots, and $T_{\text{fac}}$ for bright plages. Values for these temperatures were set according to relations found in the literature \citep{Berdyugina2005, Gondoin2008, Afram2015} and rounded to the nearest hundred. The spectra were computed at solar metallicity without alpha enrichment, and representative values of $\log g$ were picked depending on spectral type. Parameters for the stellar models can be found in Table \ref{tab:stellar_model_params}.

\begin{table}[t]
    \centering
    \begin{tabular}{l c c c c c c c c}
        \hline
        \hline
        System archetype & Spectral type & log $g$ & $T_{\text{phot}}$ (K) & $T_{\text{spot}}$ (K) & $T_{\text{fac}}$ (K) & $v \sin{i}$ (km $\text{s}^{-1}$) & $v_{\text{CB}}$ (km $\text{s}^{-1}$) & $v_{\text{syst}}$ (km $\text{s}^{-1}$) \\
        \hline
        HD 189733 & K2V & 4.5 & 5000 & 3700 & 5100 & 3.5 & 0.3 & $-2.36$ \\
        K2-18 & M2V & 5.0 & 3500 & 3000 & 3600 & 0.525 & 0.2 & 0.65 \\
        TRAPPIST-1 & M8V & 5.0 & 2600 & 2300 & 2700 & 2 & 0.2 & 0 \\
        \hline
    \end{tabular}
\caption{Stellar Model Marameters. \authorcomment1{Edited to include $v \sin{i}$, $v_{CB}$ and $v_{syst}$.}}
\label{tab:stellar_model_params}
\end{table}

For the planetary atmosphere models, the radiative transfer code petitRADTRANS was used \citep{Molliere2019}. 
Molecular volume mixing ratios were selected with two purposes: to better isolate the effect of stellar contamination on the spectral features of interest in this work (mainly $\text{H}_2$O and CO), but also to be somewhat representative of the current knowledge or models of the target archetype atmospheres (hot Jupiter, super-Earth/mini Neptune and Earth-like planet). One atmosphere model was generated for the HD 189733 and K2-18 systems. For the hot Jupiter model, absorber abundances were set to produce reasonably strong absorption features while being consistent with an extended $\text{H}_2$ envelope. The retained K2-18 model was inspired by the results of \citet{Madhusudhan2020}. For TRAPPIST-1, four models were picked, two for planet b and two for e, based on models of the evolution of atmospheres for these planets \citep{Lustig-Yaeger2019, Lincowski2018, Krissansen-Totton2018}. For TRAPPIST-1 b, we consider a model with a secondary atmosphere dominated by outgassed $\text{O}_2$ and a Venus-like model with a C$\text{O}_2$-dominated atmosphere. For TRAPPIST-1 e, the first model is of an aqua planet with an ocean surface and modern-Earth-like atmospheric abundances, and the second has abundances similar to that of the Earth in the Archean period, before significant amounts of $\text{O}_2$ were introduced to the atmosphere.

For simplicity, all model atmospheres used in this work are isothermal and well-mixed. Only a few main absorbers are included, namely $\text{H}_2$O, CO, C$\text{H}_4$, and C$\text{O}_2$, as well as Na and K in optical models. Clouds and hazes were not included in any of the models. Rayleigh scattering by $\text{H}_2$, He, $\text{O}_2$, and $\text{N}_2$ was included. The parameters and volume-mixing ratios of each atmospheric component can be found in Tables \ref{tab:planet_params} and \ref{tab:planet_VMRs}. The models are meant to be simplified versions of possible scenarios rather than rigorous, self-consistent atmosphere models.

\begin{table}[t]
    \centering
    \setlength{\tabcolsep}{3pt}
    \hskip-2.0cm
    \begin{tabular}{l c c c c c c}
        \hline
        \hline
        Planet & Model designation & $M_p$ ($M_{\text{Jup}}$) & $R_p$ ($R_{\text{Jup}}$) & $P_0$ (bar) & $P_{\text{max}}$ (bar) & $T$ (K) \\
        \hline
        HD 189733 b & Hot Jupiter & 1.13 & 1.13 & 0.01 & 100 & 1200 \\
        K2-18 b & Sub-Neptune & 0.02807 & 0.211 & 1 & 10 & 300 \\
        \multirow{2}{*}{TRAPPIST-1 b} & Outgassed $\text{O}_2$ & \multirow{2}{*}{0.0032} & \multirow{2}{*}{0.09689} & \multirow{2}{*}{0.1} & \multirow{2}{*}{10} & \multirow{2}{*}{400}  \\
         & Venus-like &  &  &  &  &  \\
        \multirow{2}{*}{TRAPPIST-1 e} & Aqua planet & \multirow{2}{*}{0.0024} & \multirow{2}{*}{0.0819} & \multirow{2}{*}{1} & \multirow{2}{*}{1} & \multirow{2}{*}{250} \\
         & Archean-Earth-like &  &  &  &  &  \\
        \hline
    \end{tabular}
    \caption{Planet Parameters Used in Atmosphere Models}
    \label{tab:planet_params}
\end{table}

\begin{table}[t]
    \centering
    \setlength{\tabcolsep}{4pt}
    \hskip-2.7cm
    \begin{tabular}{l c c c c c c c c c c}
        \hline
        \hline
         & $\text{H}_2$ & He & $\text{N}_2$ & $\text{O}_2$ & $\text{H}_2$O & CO & C$\text{O}_2$ & C$\text{H}_4$ & Na & K \\
         \hline
         Hot Jupiter & 0.86 & 0.14 & - & - & $10^{-4}$ & $10^{-4}$ & $10^{-6}$ & $10^{-6}$ & $10^{-5}$ & $10^{-6}$ \\
         Sub-Neptune & 0.85 & 0.14 & - & - & $10^{-2}$ & $10^{-6}$ & $10^{-6}$ & $10^{-8}$ & - & - \\
         Outgassed $\text{O}_2$ & $5 \times 10^{-6}$ & $5 \times 10^{-6}$ & 0.045 & 0.95 & $10^{-3}$ & $10^{-7}$ & $5 \times 10^{-3}$ & 0 & - & - \\
         Venus-like & $5 \times 10^{-6}$ & $5 \times 10^{-6}$ & 0.035 & 0 & $3 \times 10^{-5}$ & 0 & 0.965 & 0 & - & - \\
         Aqua planet & $10^{-7}$ & $10^{-7}$ & 0.8 & 0.2 & $10^{-4}$ & $10^{-7}$ & $5 \times 10^{-4}$ & $5 \times 10^{-7}$ & - & - \\
         Archean-like & $10^{-7}$ & $10^{-7}$ & 0.935 & 0 & $10^{-2}$ & $10^{-8}$ & 0.05 & $5 \times 10^{-3}$ & - & - \\
         \hline
    \end{tabular}
    \caption{Volume Mixing Ratios of Atmospheric Components for Each Atmosphere Model \authorcomment1{The sub-Neptune $\text{H}_2$ mixing ratio is indeed not a typo.} }
    \label{tab:planet_VMRs}
\end{table}

\subsection{Structure of the Models}
\label{section:structure_of_models}

To simplify the calculation of transit spectra, the model is built to work primarily with the projected 2D stellar surface. The code builds a uniform 2D Cartesian grid of $250 \times 250$ pixels and determines a circular surface with a unit radius for the star. The Python package \texttt{photutils} \citep{Bradley2019} is used to compute the overlap of the circle with every pixel on the grid. Spherical polar coordinates are assigned to every point of the stellar surface. The spherical coordinates are used to compute the radial velocity (RV) at every point given the specified rotational velocity and systemic velocity. An additional radial blueshift parameter can be specified to model an overall convective blueshift \citep{Gray2009, Shporer2011}. A value of $\mu$ (the cosine of the angle between the normal to the stellar surface and the observer's line of sight) is assigned at every point of the surface, to determine which of the specific intensity spectra corresponds to each point. The stellar models provided are binned on a relatively fine grid of $\mu$ values 
\edit1{($\Delta \mu = 0.0192$ from $\mu=1$ to $\mu=0.117$, and $\Delta \mu < 0.0001$ otherwise)}, so we have decided not to interpolate them along $\mu$.

Before introducing any temperature heterogeneity to the surface, the emergent flux of the quiet photosphere is integrated across the surface while taking into account $\mu$ and shifting the specific intensity spectra to the correct radial velocity at every pixel, to correctly account for the system velocity, stellar rotation, and convective blueshift. The purpose of this first integration is to compute the quiet emergent flux and reduce computation time when modifying the distribution of active regions.

A spot map is initialized and dark and bright areas can be added by specifying their position (in spherical coordinates) and angular radius. They are accurately projected onto the flat 2D surface. The spot map records whether a pixel is quiet or covered by a spot or facula. 
\edit1{Overlaps of spots and faculae are allowed, but pixels will record only the last active region generated on top of them.} 
Spots and faculae are added with a random distribution in position (on the sphere) and a Gaussian distribution in angular size until a specified covering fraction of the projected area is covered. In this work, we use active regions with angular radii of about 1$^{\circ}$, with a standard deviation of around 0.5$^{\circ}$. 
\edit1{The shape of active regions is implemented up to the resolution of the pixel grid; the resolution is sufficient in all cases to achieve the desired coverage fractions to a precision of 0.1\% or better (typical coverage fractions are at least 5-10 \%.)} 
A parameter controls whether active regions can be added to the transit chord region or not. In this work, we leave the transit chord intact and thus only consider unocculted spots, since we are interested in the worst-case scenarios of induced features due to stellar contamination. 

The spotted flux can then be integrated. 
To save time, this is done by going over each pixel affected by stellar activity and subtracting the difference of the\edit2{ local} spot or facula spectrum and the\edit2{ local} quiet spectrum from the\edit2{ previously integrated} quiet spectrum, with the appropriate radial velocity shifts. For models including convective blueshift, the inhibition of convective blueshift in active regions \citep{Livingston1982, Cavallini1985} is implemented simplistically, by ignoring the associated shift's contribution to the RV in active pixels when integrating the spotted flux.
\edit1{We note that, at high resolution, active regions will thus contribute differently to the stellar spectrum depending on the star's rotation and on their distribution, i.e. whether they are predominantly located east or west.}

Finally, transits are computed by introducing an opaque circle moving across the model grid. The position of this circle depends on the planet's semi-major axis and the orbital phase. All orbits are assumed to be circular. The radius of the circle depends on wavelength and corresponds\edit2{ to} the selected planetary atmosphere model, or lack of an atmosphere. 
The circle is moved at every phase and the planet radius spectrum is shifted according to the planet's changing radial velocity and the system velocity. The stellar spectrum is already shifted according to the systemic velocity from the prior steps.
At each phase of the transit, the resulting emergent flux is computed by subtracting the flux from the pixels covered by the planet. For pixels located under the planet's limb at one or more wavelengths, the \texttt{photutils} \citep{Bradley2019} package is used once more to compute the fractional overlap between the pixel and the planet. 
This is done for a number of planet radii between the minimum and maximum values, by default 1000 radii values unless specified otherwise. 
An Akima spline interpolation of the fractional overlap as a function of planet radius is then taken on the results for all pixels under the planet limb, to accelerate the computation for all the radius values in the transmission spectrum. The Akima spline is a non-smoothing type of interpolation and is chosen over a regular interpolation in order to appropriately handle the minimal and maximal points of overlap between the planet and each pixel.
Most of the transit time series spectra are computed at a spectral resolution of $R=100,000$, unless mentioned otherwise. A few models are computed at $R=200,000$ to test for the effect of even higher resolutions that could eventually be reached with future instruments.

Photon noise can then be added to the data to approximately simulate observations. The transit time series is first normalized with the 2MASS J-band response function and the assumed J-band magnitude of the model host star. We take the HD 189733, K2-18, and TRAPPIST-1 values from the 2MASS catalog\edit1{: 6.07, 9.76, and 11.35, respectively} \citep{Cutri2003}. Finally, flux values are converted to photon counts using assumed instrumental parameters: collecting area, throughput \edit1{(taken as a constant across the wavelength domain)}, exposure time, and number of visits. Other sources of noise such as instrumental readout noise and telluric transmission are not considered in this analysis. Our motivation is to look at "optimal" detections of worst-case scenarios, to estimate what kind of signal-to-noise ratios (S/N) one should aim to achieve for each target.

As an aside, it is also possible to make the star (i.e. the spots and faculae) rotate across a full rotation period, and to record the resulting spectrophotometric curve. In order to obtain an accurate estimate of stellar variability, active regions are also generated on the hidden hemisphere of the star, with similar coverage fractions. This allows active regions to rotate in and out of the visible hemisphere. The position of active regions is updated with every phase increment of the star's rotation, and the emergent flux is integrated at every phase across the full near infrared wavelength domain \edit1{to obtain photometric measurements}. This can be used to try to relate spot/faculae covering fractions and photometric variability as a possible diagnostic tool for stellar activity, as mentioned in \citet{Rackham2017} and subsequent papers. The current implementation is not computationally efficient, prohibiting extensive testing for the moment.

A few major assumptions are made about the geometry of the modeled systems. Stellar surface radial velocities are calculated assuming solid body rotation (i.e. no differential rotation). Spin-orbit misalignment is left at 0 for all systems. However, orbital inclination is not forced to be $90^{\circ}$: all planet models are given low impact parameters.

\subsection{Data Analysis}
\label{section:data_analysis}

The resulting transit spectra time series are analyzed at low and high resolution. The spectra are degraded to a lower resolution by convolving them with a Gaussian kernel of appropriate width and binning them to the new wavelength grid. Resampling is done before performing the analysis, when required.

\pagebreak

\subsubsection{Light-curve Fitting and Low-resolution Analysis}

At low resolution, the transmission spectra are obtained by fitting the normalized transit light curve for every wavelength bin. 
\edit1{Photon noise can be added to the spectra before fitting, after the convolution to the instrument resolution.}
The light curves are normalized at all wavelengths simply by dividing by the average out-of-transit flux. The Python packages \texttt{batman} \citep{Kreidberg2015} and \texttt{emcee} \citep{Foreman-Mackey2013} are used to perform a Monte Carlo Markov Chain (MCMC) fit with the effective planet radius, limb darkening coefficients, and a noise parameter as the free parameters. A single quadratic limb-darkening law is chosen for the fit, assuming a homogeneous photosphere. The time since mid-transit, orbital semi-major axis, eccentricity, orbital period, inclination, and longitude of periastron are all fixed. We take 60 walkers for 3000 steps, with 500 steps discarded as burn-in.

The resulting spectra are visually compared for each case to identify similarities and differences in the spectral features due to contamination and the planet atmosphere, as well as how stellar contamination modifies an existing signal from an atmosphere. We look for apparent degeneracies in the appearance of spectral features. We can also observe any overall radius offsets associated with stellar contamination, as well as slopes.

\subsubsection{Cross-correlation and High-resolution Analysis}
\label{section:high_R_analysis}

For the high-resolution analysis, photon noise is added to the observed spectra $F_i$ before any further manipulation of the spectra in order to estimate which kinds of telescopes and instruments would be required to achieve detections on the modeled targets. 
\edit2{The only manipulation done to the low-resolution spectra before adding noise is the convolution to instrumental resolution.} \authorcomment1{this was added for clarification} \edit2{The manipulations we refer to in the high-resolution case are the mean subtraction and normalization operations.}
Two telescope classes are considered: one like CFHT with collecting area of about 8 $\text{m}^2$, and an extremely large telescope (ELT) like the E-ELT with collecting area of 978 $\text{m}^2$. A master out-of-transit spectrum $F_{\text{OOT}}$ is obtained by averaging all of the out-of-transit spectra and each observed spectrum during transit is divided by $F_{\text{OOT}}$. 

A cross-correlation analysis is performed on the resulting transmission spectra. The goal is to obtain a 2D correlation map, i.e., to correlate the spectra with a line list ($\text{H}_2$O or CO) or a reference spectrum as a function of the systemic velocity $v_{\text{syst}}$ and the planet's orbital velocity semi-amplitude $K_p$. 

For the correlation, the spectra are normalized to a mean of 0 and self-correlation of 1 by applying equations \ref{eq:CC_subtract_mean} and \ref{eq:CC_normalize} to the transmission spectra $S_i = 1-F_i/F_{\text{OOT}}$, where $i$ corresponds to the individual observations during the transit itself. This normalizes the perfect correlation value to 1 for an individual CCF. The spectra used to compute the correlation maps are denoted $f_i$.

\begin{equation}
    \Bar{S_i} = S_i - \frac{1}{n_{\lambda}} \sum_{\lambda} S_i (\lambda)
    \label{eq:CC_subtract_mean}
\end{equation}

\begin{equation}
    f_i = \frac{\Bar{S_i}}{(\sum_{\lambda} \Bar{S_i}(\lambda)^2)^{1/2}}
    \label{eq:CC_normalize}
\end{equation}

To obtain a correlation map, the correlation with the desired line list or reference spectrum is computed for each observation during transit and added over phases for each combination of $v_{\text{syst}}$ and $K_p$ according to equation \ref{eq:CC_map}. In this work, line lists are used directly \edit1{as cross correlation-templates to detect specific molecules}. The observed spectra are deshifted according to $v_{syst}$ to be brought back to the observer rest frame, which we denote $f_i (\lambda_{v_{\text{syst}}})$. The line list or reference spectrum is denoted $g$. For each observation $i$, the reference spectrum is shifted to the radial velocity the planet would have in that instant in the star's rest frame. This shift depends on $K_p$ as well as the time during transit, thus it is denoted by writing $g(\lambda_{K_p,i})$. $K_p$ is computed assuming a circular orbit.
Planetary signals are expected to produce a correlation peak around the true values of $K_p$ and $v_{\text{syst}}$ of the model, and signals due to stellar activity contamination are supposed to appear around $K_p = 0$ and the true $v_{\text{syst}}$, assuming they are distributed roughly uniformly on the stellar surface.
\edit2{Deviations from this can be due to nonuniform active regions distributions as well as imperfect correction of the RME.}

\begin{equation}
    \text{CCF}(v_{syst}, K_p) = \sum_{i} \sum_{\lambda} f_i (\lambda_{v_{syst}}) g(\lambda_{K_p,i})
    \label{eq:CC_map}
\end{equation}

\begin{figure}[ht]
    \centering
    \includegraphics[width=\textwidth]{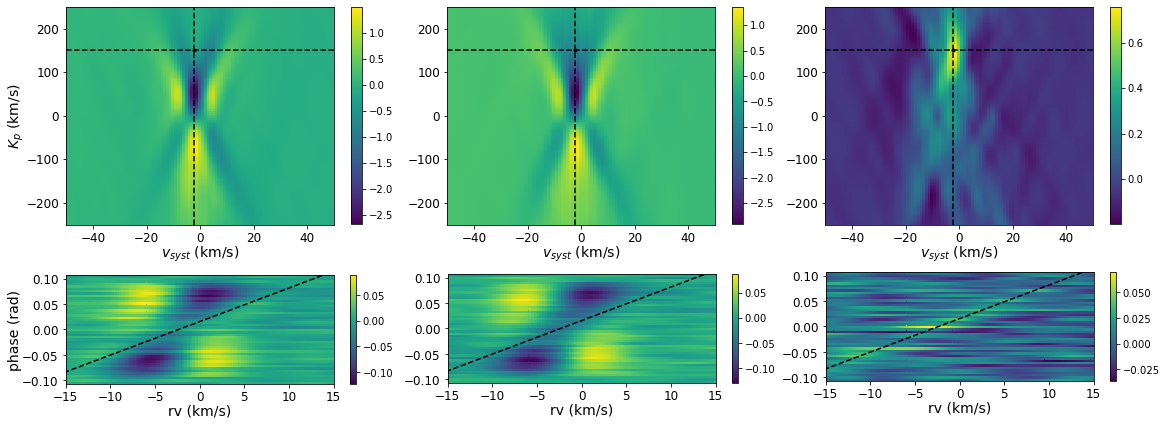}
    \caption{Example of the RME correction process for the HD 189733 b model with the hot Jupiter atmosphere, but without stellar activity. Cross correlation is taken with a CO line list \citep{Rothman2010} at 1200 K across the 2.29 $\mu$m CO band. The dashed lines indicate the true systemic and planetary RVs. \textit{Left panel:} Correlation map (\textit{top}) and CCF for individual spectra during transit (\textit{bottom}) before applying any correction. \textit{Middle panel:} Same, but for a bare rock planet. This includes only the correlation signal due to the occultation of the stellar surface. \textit{Right panel:} Same again, after subtracting the RME correlation (middle panel) from the uncorrected correlation (left panel). The planet signal is now clearly visible, especially in the correlation map. }
    \label{fig:HD_189_correction_process}
\end{figure}

In some cases, the RME affects stellar molecular lines that are used in correlation, leading to spurious signals in the correlation map that can dominate over a true planetary signal. One way this can be corrected for is by modeling how the RME affects the correlation signal, i.e. by computing a transit for a homogeneous photosphere and a planet without an atmosphere and performing the same correlation analysis. This RME correlation signal can subsequently be subtracted from the correlations of the transit models we wish to analyze. In principle, this allows us to recover planet signals as well as any stationary contamination signal. This process is illustrated in Figure~\ref{fig:HD_189_correction_process}.

The necessity of correcting for the RME has already been pointed out in other efforts \citep[e.g.][]{Brogi2016, Casasayas2019} and we use similar methods to apply our correction. The main downside of using this type of correction is that it requires either prior knowledge of the star's $v \sin{i}$ and spin orbit misalignment $\lambda$, or a separate fit for these parameters in order to model the stellar surface.

The correlation maps in this work will be plotted after normalization by a standard deviation $\sigma$. We choose to exclude the main peak from the calculation of $\sigma$. Thus, to obtain $\sigma$ for a system and a number of visits, we compute multiple maps with different realizations of photon noise. The maps are then subtracted from one another to eliminate the main peak/feature and conserve only random noise. The correlation noise $\sigma$ can then be estimated by taking the standard deviation of the resulting maps.
For RME-corrected maps, we use the same subtraction procedure to compute $\sigma$ as we use to obtain the actual maps, so we use this $\sigma$ itself as is.
For uncorrected maps, we use $\sigma / \sqrt{2}$ to take into account the effect of the subtraction on the maps used to compute $\sigma$.


\section{Results}
\label{sec:Results}

For all model classes, we look at what we could call "worst-case scenarios" of stellar contamination, to determine how much activity is required in order to significantly challenge planetary atmosphere detections. Therefore, we consider only unocculted active regions, such that the strength of the transit light source effect is maximized. This assumption would be true in the case of a star with polar active regions and a planet with a low impact parameter, for example, but this is likely not the case for most target systems for transit spectroscopy.

We test various possible spot and faculae coverage fraction combinations: models with only spots, with only faculae, and with combinations of both. There is an important degeneracy in terms of the effect these combinations of covering fractions have on the observed spectral signatures. Since we use fixed temperature contrasts between the active regions and the photosphere, it appears that the absolute values of the covering fractions are not always as important as the ratio between spot and faculae area. Naturally, it also follows that the results shown here are limited by the assumptions that were made about active region temperatures.

We also combine the effect of active regions with planetary atmospheres to see if we can introduce new molecular features in a transmission spectrum, accentuate a faint feature, leading to an overestimate of the molecular abundance, or even mute a feature.

For each model class, we present our most interesting results, first at low resolution and then at high resolution.

\subsection{Some Photometric Results}

\begin{table}[ht]
    \centering
    \setlength{\tabcolsep}{3pt}
    \hskip-2.0cm
    \begin{tabular}{l c c c}
        \hline
        \hline
        Model system & Spot coverage (\%) & Faculae coverage (\%) & Variability amplitude (\%) \\
        \hline
        \multirow{3}{*}{HD 189733} & 5 & 0 & \edit1{0.60} \\
         & 10 & 0 & \edit1{1.1} \\
         & 5 & 15 & \edit1{0.63} \\
        \hline
        \multirow{2}{*}{K2-18} & 10 & 0 & \edit1{0.63} \\
         & 20 & 0 & \edit1{0.95}  \\
        \hline
        \multirow{2}{*}{TRAPPIST-1} & 10 & 0 & \edit1{0.93} \\
         & 20 & 25 & \edit1{1.2} \\
        \hline
    \end{tabular}
    \caption{Photometric Variability Amplitude Across a Full Stellar Rotation. \authorcomment1{Values updated to reflect the new computations.}}
    \label{tab:photometric_variation}
\end{table}

\begin{figure}[t]
    \centering
    \includegraphics[width=\textwidth]{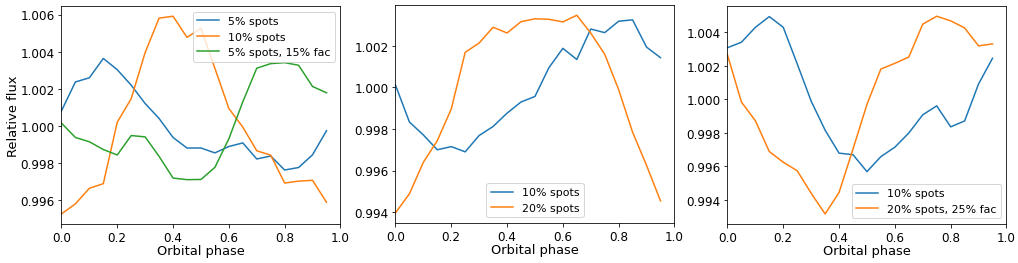}
    \caption{\edit1{Examples of photometric variability curves for the active region coverage fractions in Table~\ref{tab:photometric_variation}. \textit{Left}: HD 189733. \textit{Center}: K2-18. \textit{Right}: TRAPPIST-1. } \authorcomment1{New figure.}}
    \label{fig:variability_curves}
\end{figure}

A few tests were performed to determine the photometric variation associated with some of the models used in this study, using the method described at the end of section~\ref{section:structure_of_models}. The results are presented in Table~\ref{tab:photometric_variation}. They lie in the range of what can be expected from late-type stars \citep[][]{Newton2016}. We note that all the computed variation curves were \edit1{relatively} smooth and sinusoidal\edit1{, as shown in Figure~\ref{fig:variability_curves}}. These values are only examples of the variability such spot and faculae covering fractions can produce, and therefore do not represent the full range of variability amplitudes that can be caused by different active region distributions. For a more detailed study on this topic, we refer to~\citet{Rackham2018,Rackham2019}. \authorcomment1{New computations of photometric variability were conducted upon realizing that there were previously some an issue with the uniformity of generated spot positions. This issue did not affect the computation of transit spectra, only that of photometric variability.}

\subsection{HD 189733 Analog System}

The models based on the HD 189733 system are different from the next two. Because of the higher stellar temperature, stellar contamination does not affect water features. Instead, the presence of CO in the photosphere and active regions as well as continuum effects are the main components of stellar contamination in our models.

\subsubsection{Low Resolution}

\begin{figure}[t]
    \centering
    \includegraphics[width=0.65\textwidth]{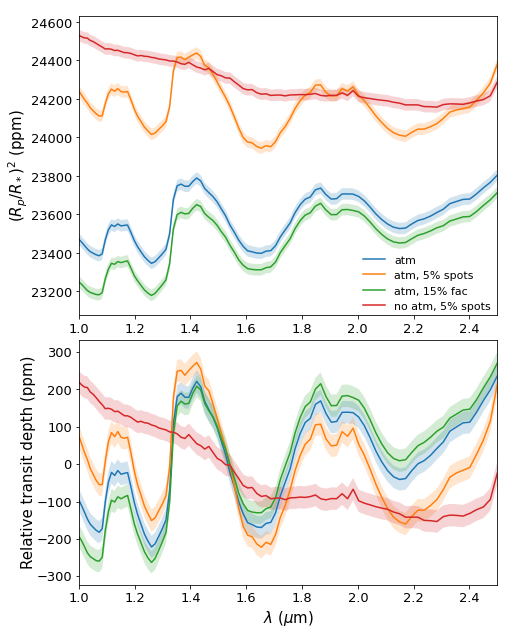}
    \caption{Comparison of near infrared low-resolution ($R=100$) transit spectra for the HD 189733 b-type models.\edit1{ Uncertainty estimates are indicated as for 3 JWST transits, binned to $R=100$.} \textit{Top:} the atmosphere model transit (blue) is compared to a pure stellar contamination transit with 5\% dark spots (red), an atmosphere transit with 5\% spots (orange), and an atmosphere transit with 15\% bright faculae (green). \textit{Bottom:} Relative transit depth of three of the models (same colors). \authorcomment1{New figure including 1-$\sigma$ uncertainties.}}
    \label{fig:HD_189_low_R_IR}
\end{figure}

\begin{figure}[t]
    \centering
    \includegraphics[width=0.65\textwidth]{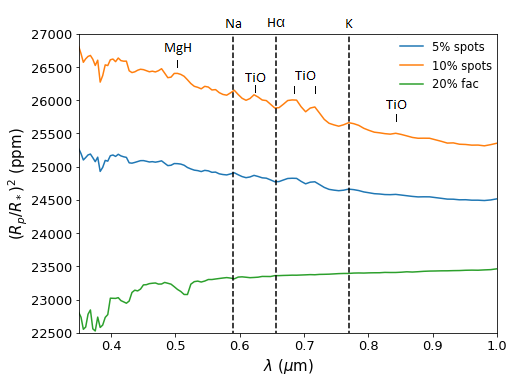}
    \caption{Comparison of optical low-resolution ($R=100$) transit spectra for the HD 189733 b-type models.\edit1{ Shaded uncertainty regions are omitted here, since they are very small.} The plot includes only stellar contamination transit spectra: 5\% spots (blue), 10\% spots (orange), and 20\% faculae (green). \edit1{The positions of some key atomic and molecular features are highlighted.} \authorcomment1{New figure highlighting some likely contamination features.}}
    \label{fig:HD_189_low_R_optical}
\end{figure}

In the low-resolution regime ($R=100$), stellar contamination does not appear to affect the main atmospheric features present in the near infrared transmission spectra, as shown in Figure \ref{fig:HD_189_low_R_IR}. It does, however, introduce a "slope" to the spectra as well as an offset in the overall inferred radius. The slope is present in both the optical and near infrared domains, but is overall more pronounced in the optical, as can be seen in Figures \ref{fig:HD_189_low_R_IR} and \ref{fig:HD_189_low_R_optical}.

Both the slope and offset should be due to the difference in overall temperature between the transit chord and the rest of the stellar surface, i.e. if the transit chord is on average brighter (less dark spots) than the rest of the surface, then the transit will appear deeper, and vice versa. This effect has already been pointed out by other teams in previous efforts, \citep[e.g.][]{Rackham2019}. While these effects may not be relevant to molecular detections, they can still very much affect the characterization of an atmosphere by mimicking the contribution of Rayleigh scattering or hazes if unaccounted for \citep[e.g.][]{Pont2008,Sing2011}.

\edit1{In the optical contamination spectra (Figure~\ref{fig:HD_189_low_R_optical}), we observe relatively weak features that appear to be related to absorption by Na (0.589 $\mu$m), K (0.77 $\mu$m), TiO \citep{Sedaghati2017, McKemmish2019}, and MgH \citep{GharibNezhad2013}, as well as possible H$\alpha$ (0.656~$\mu$m) emission. These are highlighted in Figure~\ref{fig:HD_189_low_R_optical}. Features of other metal oxides or hydrides may also be affected by stellar contamination.}

\subsubsection{High Resolution}

Even though stellar contamination did not appear to introduce any major issues in low-resolution spectra when it comes to molecular detections, we still wish to investigate the effects of contamination on high-resolution observations. We consider transit spectra with pure photon noise as for visits with a 3.6 m class telescope such as CFHT. We take two visits when looking at $\text{H}_2$O and three when looking at CO.
\edit1{We keep the same active region distribution during visits; "visits" are used here strictly as a way to evaluate noise levels.}

At very high resolution ($R=100,000$), the effect of stellar contamination depends on which molecules are being studied and which molecular lines appear in the stellar spectrum. In the case of an early-K dwarf such as HD 189733, water should not be present in the stellar spectrum. Therefore, the RME should not pose a problem when trying to detect water in the atmosphere of a planet around such a star. CO lines, on the other hand, do appear in the spectra of these stars. If the RME and the effect of convective blueshift are not properly accounted for, they will affect attempts at detecting CO in an exoplanet atmosphere. 

For mixed transit models (atmosphere + stellar contamination), if contamination induces features for the molecule of interest, recovering the planet signal may require additional steps, since the spot and planet signals are superposed. Furthermore, the contamination signal can be strong enough to mask the planet signal. As a solution, we will use the correction method described at the end of section \ref{section:high_R_analysis}. 

\begin{figure}[ht]
    \centering
    \includegraphics[width=\textwidth]{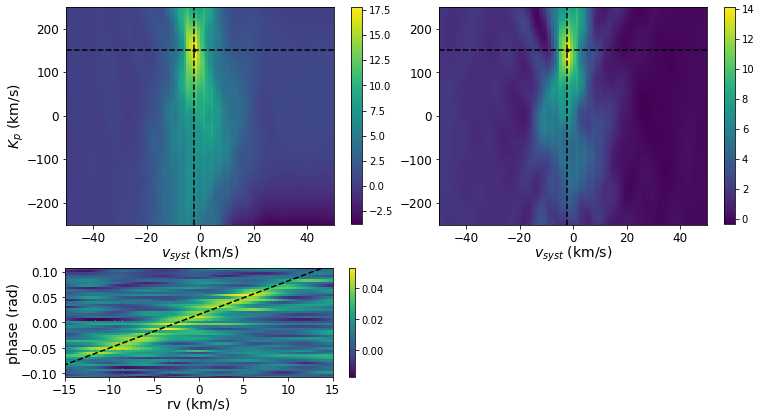}
    \caption{\textit{Top:} correlation map\edit1{s} (normalized by standard deviation) for a transit model including the hot Jupiter atmosphere as well as 5\% spot coverage on the host star. The true position of the planet signal is indicated by the black dot. \edit1{\textit{Left}:} \edit1{correlation} with a water line list \citep{Polyansky2018} at 1200 K across the 1.4 $\mu$m water band. No RME/CLV correction was applied. Correlation values are normalized by the standard deviation across the map, not including the peak. Photon noise was added as if the observations were taken during two transits from SPIRou/CFHT. \textit{Bottom:} corresponding CCF for individual spectra during transit, with the dashed curve indicating the true planet RV. \edit1{\textit{Right}: correlation with a CO line list \citep{Rothman2010} at 1200 K across the 2.29 $\mu$m CO band. The correction method (as shown in Figure~\ref{fig:HD_189_correction_process}) was applied. Photon noise was added as for three SPIRou/CFHT visits.} \authorcomment1{Combination of previously Figures 4 and 5.}}
    \label{fig:HD_189_corr}
\end{figure}

In the case of water, we are able to recover the cross-correlation signal easily without performing any correction for stellar signals, as expected. The correlation map of a transit with a water line list \citep{Polyansky2018} is shown in Figure \ref{fig:HD_189_corr}, with a clear peak centered at the true planet orbital parameters.

In the case of CO, correcting for stellar effects is necessary in order to retrieve a planetary signal. 
As shown in the left and middle panels of Figure \ref{fig:HD_189_correction_process}, there are strong signals in the correlation map of a transit and a CO line list \citep{Rothman2010} that are due to the RME and CLV, i.e. effects due to variations across the stellar surface. There are two main peaks: one strong negative peak at the true $v_{\text{syst}}$ and around $K_p=50$ $\text{km s}^{-1}$, and a slightly weaker positive peak mirroring it at $K_p=-50$ $\text{km s}^{-1}$. 
The negative peak is associated with the narrow bump in stellar line profiles created by the planet occulting a small region of the star, and the positive peak comes from the change in the overall line profiles.
For example, in the first part of the transit (bottom of the lower panels in Figure~\ref{fig:HD_189_correction_process}), when the planet covers a blueshifted part of the star, the blueshifted bump in the spectral lines produces the negative peaks. At the same time, the spectral line profile gains an overall redshift, which produces the positive peaks. In the middle of transit (at phase 0), there is no effective shift, since only the center of the lines is affected. In the final half of the transit, the shifts are reversed.
The value of $K_p = \pm 50$ $\text{km s}^{-1}$ is related to the $v \sin i$ of the stellar model and the peaks are centered at $v_{\text{syst}}$ because they are stellar in origin. We note that these peaks are present whether the star has active regions or not. Fortunately, it appears that we are able to recover the planet signal by applying the correction method described in section \ref{section:high_R_analysis} and Figure~\ref{fig:HD_189_correction_process}, as shown in Figure~\ref{fig:HD_189_corr}. In this case, the transit model also included 5\% spot coverage on the star, but the signal from these active regions is not strong enough \edit1{to} affect the atmospheric detection.

\subsection{K2-18 Analog System}

In the case of the K2-18 class of models and late-type stars in general, our main concern and focus when it comes to stellar contamination is water features. We note that other molecules such as CO or C$\text{O}_2$ could be involved, but for the scope of this work, we choose to focus on water.

\subsubsection{Low Resolution}
\label{section:K2-18_low_resolution}

\begin{figure}[t]
    \centering
    \includegraphics[width=0.65\textwidth]{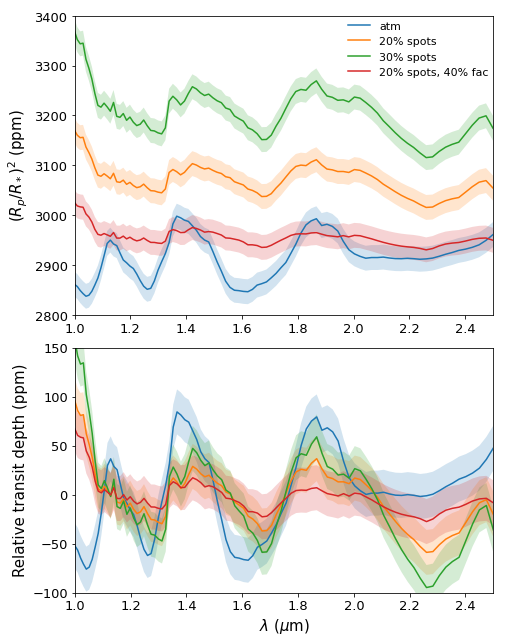}
    \caption{Comparison of the low-resolution ($R=100$) atmospheric transit spectrum and pure stellar contamination transits for the K2-18 type model.\edit1{ Uncertainty estimates are indicated as for 10 JWST transits, binned to $R=100$.} \textit{Top:} pictured are the pure atmospheric transit (blue), a 20\% spot coverage transit (orange), a 30\% spots transit (green), and a transit with 20\% spots and 40\% faculae (red). \textit{Bottom:} Relative transit depth of all the models (same colors). \authorcomment1{New figure including 1-$\sigma$ uncertainties.}}
    \label{fig:K2-18_atm_vs_pure_contamination}
\end{figure}

At $R=100$, the amplitude of contamination features is relatively weak even for large coverage fractions of strictly unocculted spots, much weaker than we will see in the TRAPPIST-1 models, although the "worst-case" scenarios we consider do in fact reach amplitudes that are almost comparable to that of a temperate sub-Neptune atmosphere, as shown in Figure \ref{fig:K2-18_atm_vs_pure_contamination}. If we look at the 1.4 $\mu$m water band, the amplitude of the planetary absorption is of about 150 ppm, while the strongest contamination amplitude (from the 30\% spots model) reaches almost 100 ppm. Contamination features are made weaker when the dark spots are balanced with large covering fractions of bright faculae, around 25 to 40\% for example, in which case the amplitude of the 1.4 $\mu$m water band from contamination drops to about 25 ppm.

We note that, for a different type of planet --- for example a smaller planet with a high molecular weight atmosphere --- the atmospheric features and contamination features could be much more similar. They would both be significantly more challenging to detect for such a system, however.

One key potentially helpful difference between the planet and star water features is the water absorption band around 1.1-1.2 $\mu$m, which is present in planet transmission spectra with an amplitude of about 100 ppm, but is essentially absent in contamination spectra. This is due to the band being strong in water's absorption spectrum at 300 K (planet), but essentially absent at 3000 K (starspots). We can also notice that the general shape of the absorption features in the contamination spectra is different from the shape of the planetary features: they are wider, with a flatter peak.
These differences in the water absorption spectrum at 300 K and 3000 K could prove helpful, provided the features can be constrained well enough by observations. 
Finally, the slope of the spectra between 1.0 and 1.4 $\mu$m is also different, with the atmospheric spectrum being generally flat outside of absorption bands and the contamination spectra having a sharp slope, particularly between 1.0 and 1.1 $\mu$m.

\begin{figure}[t]
    \centering
    \includegraphics[width=0.65\textwidth]{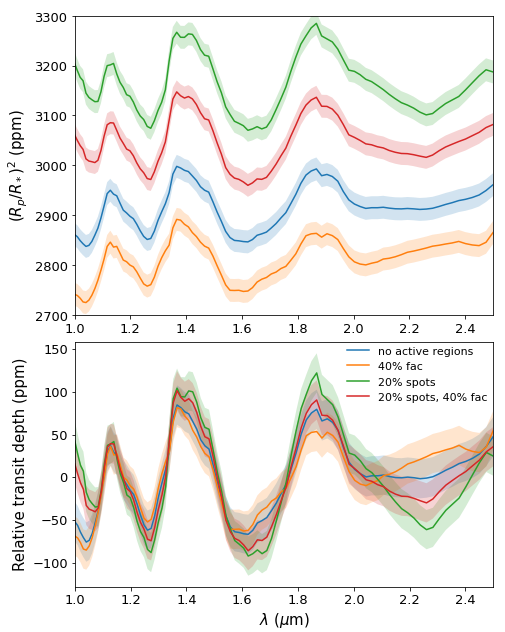}
    \caption{Comparison of the K2-18 type model atmosphere transits with and without stellar contamination.\edit1{ Uncertainty estimates are indicated as for 10 JWST transits, binned to $R=100$.} \textit{Top:} the uncontaminated transit (blue) is plotted along with transits with 40\% faculae (orange), 20\% spots (green), and 20\% spots with 40\% faculae (red). \textit{Bottom:} Relative transit depth for the same models (same colors). \authorcomment1{New figure including 1-$\sigma$ uncertainties.}}
    \label{fig:K2-18_mixed_models}
\end{figure}

In mixed models (atmosphere and stellar contamination) the shape of water signatures appears mostly unaffected except past 2.1 $\mu$m, as shown in Figure \ref{fig:K2-18_mixed_models}. However, their amplitude can be modulated by over 50 ppm in the case of 20\% spots. While an atmospheric detection can be achieved with reasonable confidence even in the presence of stellar contamination -- especially if there is no reason to think the star should be very active -- the effect on the amplitude of features could certainly affect a retrieval of atmospheric abundances. The planet radius measurement can be affected, too, by as much as 6-7\%, which can in turn strongly bias density measurements.

From previous observation campaigns \citep[e.g.][]{Benneke2019}, we know it is possible to reach precision levels down to at least about 30 ppm in each \edit1{0.1 $\mu$m wide} wavelength channel of HST/WFC3 on a target like K2-18 b, provided sufficient transits. Therefore, the signatures of strong stellar contamination (over 20\% spots) can affect transmission spectra above noise levels. The most likely impact on atmospheric characterization efforts would be systematic errors on retrieved water abundances. Entirely false-positive detections appear possible for such high activity levels, highlighting the need for at least some level of constraints on the star's level of activity. See section~\ref{sec:Discussion} for a few more comments on the existing water detections for K2-18 b.

\authorcomment1{Upon reading another upcoming paper regarding the problem of stellar contamination and the water detection in K2-18 b \citep{Barclay2021}, we decided to delve further into the optical, since our available spectra allow it. We see that including the K2 photometric data point can help discriminate between stellar contamination and atmosphere models. We included this, along with an additional figure, in the discussion.}

\subsubsection{High Resolution}

We first note that the orbital parameters taken from the true K2-18 system include a wide semi-major axis and a long period, resulting in a small change of planet orbital velocity during transit compared to the usual targets for high-resolution transit spectroscopy (e.g. planets in close-in orbits). 
\edit1{For example, our K2-18 b model undergoes a total radial velocity variation of about 1.2 $\text{km s}^{-1}$ across the full transit, compared to 32 $\text{km s}^{-1}$ for our HD 189733 b model.}
We therefore perform correlations using the normal system parameters as well as an artificially boosted $K_p$. The boost is obtained only by adjusting the semi-major axis and orbital period in accordance with Kepler's third law. Obviously, the temperature of the atmosphere model used becomes inconsistent with the new orbit, but this should still provide information about the detectability of an atmosphere in the case where the planet shows a larger radial velocity variation during transit. We also test whether extremely high spectral resolution ($R=200,000$) can provide better constraints on the correlation analysis for the normal orbit.

\begin{figure}[t]
    \centering
    \includegraphics[width=0.6\textwidth]{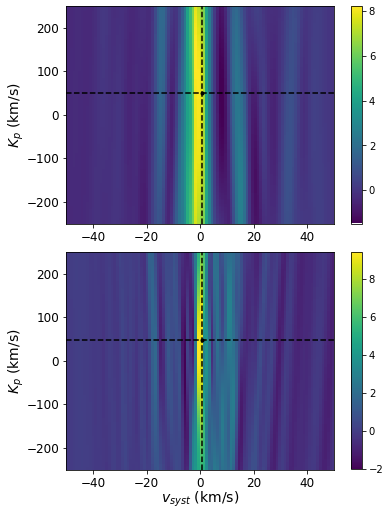}
    \caption{Correlation maps (normalized by standard deviation) for a transit model with a sub-Neptune atmosphere and 10\% spot coverage on the host star, correlated with a water line list \citep{Polyansky2018} at 300 K. The true planetary RV (non-boosted) was used. The correction for the RME/CLV was applied. \textit{Top:} $R=100,000$, with photon noise added to the spectra as if they had been taken from 30 SPIRou/CFHT visits. \textit{Bottom:} $R=200,000$, with photon noise as for 30 SPIRou/CFHT visits.}
    \label{fig:K2-18_corr_normal}
\end{figure}

\begin{figure}[t]
    \centering
    \includegraphics[width=0.6\textwidth]{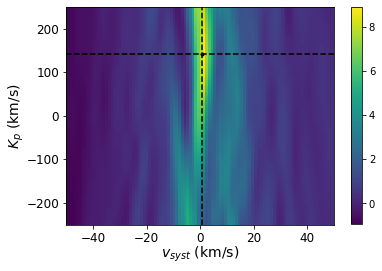}
    \caption{Same as Figure \ref{fig:K2-18_corr_normal}, but in this case the orbital velocity of the planet was artificially boosted to $K_p=141$ $\text{km s}^{-1}$ and 30 SPIRou/CFHT visits were used for the photon noise. The $R=100,000$ model was used here.}
    \label{fig:K2-18_corr_boosted}
\end{figure}

We present correlation maps for mixed transit models (sub-Neptune atmosphere with 10\% spot coverage on the star) in the normal and boosted cases in Figures \ref{fig:K2-18_corr_normal} and \ref{fig:K2-18_corr_boosted}, respectively. Once again, pure photon noise is added as for 3.6 m telescope visits, this time using 30 visits. We look only at the signal from water lines in the water band centered around $1.4 \mu$m. The effects of RME and CLV appear to be negligible in this case, or are difficult to see on the maps. This is explained by the slower stellar rotation of the K2-18 model ($v \sin i = 0.5$ $\text{km s}^{-1}$ compared to 3.5 $\text{km s}^{-1}$ in the previous models), as well as the very small size of the model planet in this case (transit depth of 2800 ppm against 24,000 ppm for the HD 189733 b model). Stellar line profiles are therefore much less affected by the transiting planet. Nevertheless, in both cases, we only consider the maps for which the correction procedure was applied.

While the peak in the boosted case (Figure \ref{fig:K2-18_corr_boosted}) corresponds without a doubt to an atmospheric detection at the expected $K_p$ and $v_{\text{syst}}$, the signal in the non-boosted case (Figure \ref{fig:K2-18_corr_normal}) is very poorly constrained with respect to $K_p$, especially at $R=100,000$, covering almost the entire $K_p$ range used for calculation and thus making an unambiguous detection difficult. We do note, however, that at extremely high resolution ($R=200,000$), the signal is better constrained with respect to $K_p$, even in the non-boosted case.

\begin{figure}[t]
    \centering
    \includegraphics[width=0.6\textwidth]{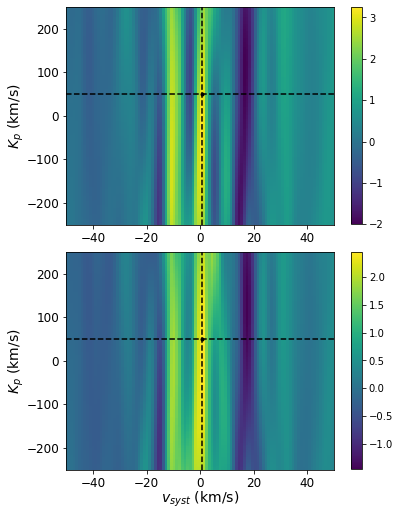}
    \caption{Correlation map (normalized by standard deviation) for a transit model ($R=100,000$) with only 10\% spot coverage (no planetary atmosphere) on the K2-18--like host star, correlated with a water line list \citep{Polyansky2018} at 300K. \edit1{\textit{Top:} Before correcting for the RME/CLV.} \edit1{\textit{Bottom:}} After the RME/CLV correction was applied\edit1{.} Photon noise was added for 30 SPIRou/CFHT visits\edit1{ in both cases}.}
    \label{fig:K2-18_corr_spots}
\end{figure}

In comparison, we show in Figure \ref{fig:K2-18_corr_spots} the correlation map for a signal due only to a 10\% spot coverage on the star (no planetary atmosphere). The long vertical positive peak at the true $v_{\text{syst}}$ present in Figure \ref{fig:K2-18_corr_normal} is absent here, confirming that the source of said signal is indeed the planetary atmosphere. \edit1{The features present in the map in Figure~\ref{fig:K2-18_corr_spots} are weak and generally inconsistent between different noise calculations. Therefore, we do not consider them to be a detection.}

\authorcomment1{The shifts observed in the previous version of Figure 10 were most likely due to the fact that the signal is not fully detected in the spots-only case. }

\subsection{TRAPPIST-1 Analog System}

\subsubsection{Low Resolution}

The features induced by 10--20 \% dark spots are comparable with most atmosphere models with prominent water features, so we keep this as an upper bound for the results we present. Lower coverages with weaker signal can still affect atmosphere characterization and abundance retrievals, as well \edit1{as} planet radius/density measurements due to flat offsets to the transit depth.

\begin{figure}[ht]
    \centering
    \includegraphics[width=0.65\textwidth]{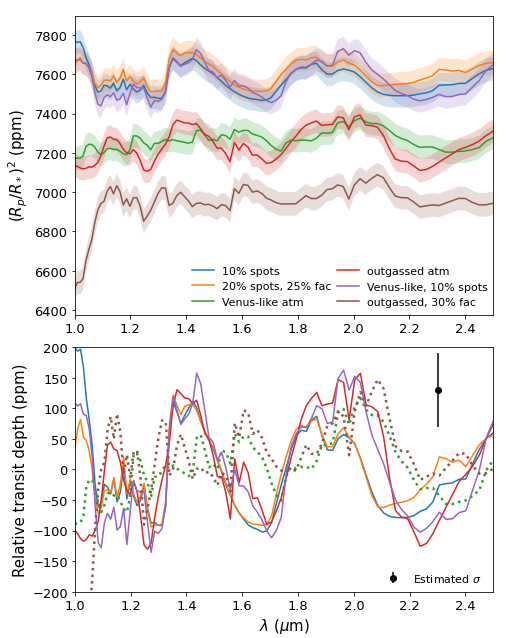}
    \caption{Comparison of multiple low-resolution ($R=100$) transit spectra for different TRAPPIST-1 b models.\edit1{ Uncertainty estimates are indicated as for 10 JWST transits, binned to $R=100$.} \textit{Top:} the plot includes the uncontaminated outgassed atmosphere model (red), the uncontaminated Venus-like model (green), a pure contamination transit with 10\% spot coverage (blue), a pure contamination transit with 20\% spots and 25\% faculae (orange), the Venus-like model with 10\% spots (purple), and the outgassed model with 30\% faculae (brown). \textit{Bottom:} Relative transit depth for four of these models (same colors). \authorcomment1{New figure including 1-$\sigma$ uncertainties.}}
    \label{fig:TRAPPIST_b_low_R}
\end{figure}

A comparison of various models for a planet like TRAPPIST-1 b is shown in Figure \ref{fig:TRAPPIST_b_low_R}. Stellar contamination poses more issues here than in previous models. We observe that the outgassed $\text{O}_2$ atmosphere model has water features similar to those of the pure stellar contamination models (10\% spots as well as 20\% spots with 25\% faculae), although there is a significant radius offset of about 400 ppm in transit depth between the spectra. The Venus-like model can be made to exhibit strong water signatures by adding 10\% spot coverage to the star, at which point it closely matches the outgassed $\text{O}_2$ atmosphere. The strongest similarity between all these models can be seen clearly when they are offset to the same level in the bottom panel of Figure \ref{fig:TRAPPIST_b_low_R}: the 1.4 $\mu$m water band of all four models (outgassed, Venus-like, and the two pure contamination models) overlap almost perfectly and have the same amplitude of about 200 ppm. The differences between the planet and contamination models around 2.0 $\mu$m are due to C$\text{O}_2$ absorption. The only way to establish the planetary origin of water absorption in this case would be to achieve sufficient S/N to clearly detect the absorption band at 1.1 $\mu$m (about 150 ppm in amplitude) where a difference was also observed in the K2-18 models. Even then, it would be extremely challenging to characterize the atmosphere. Finally, we observe that the \edit1{1.4 and 1.9 $\mu$m} water absorption bands of the outgassed model are almost completely muted by adding 30\% faculae coverage, in which case it would be \edit1{difficult} to recover a water detection.
\edit1{Provided sufficient S/N, a retrieval algorithm that includes a treatment of stellar activity may be able to detect water via the 1.1 $\mu$m band.}

\begin{figure}[ht]
    \centering
    \includegraphics[width=0.67\textwidth]{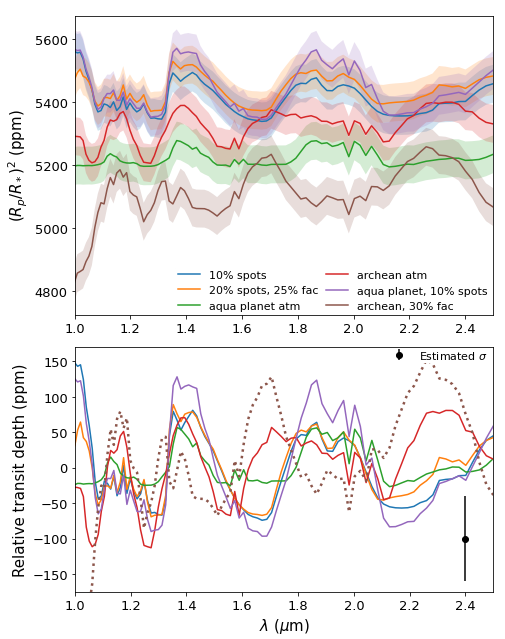}
    \caption{Comparison of multiple low-resolution ($R=100$) transit spectra for different TRAPPIST-1 e models.\edit1{ Uncertainty estimates are indicated as for 10 JWST transits, binned to $R=100$.} \textit{Top:} the plot includes the uncontaminated Archean-Earth--like atmosphere model (red), the uncontaminated aqua planet model (green), a pure contamination transit with 10\% spot coverage (blue), a pure contamination transit with 20\% spots and 25\% faculae (orange), the aqua planet with 10\% spots (purple), and the Archean-like model with 30\% faculae (brown). \textit{Bottom:} Relative transit depth for four of these models (same colors). \authorcomment1{New figure including 1-$\sigma$ uncertainties.}}
    \label{fig:TRAPPIST_e_low_R}
\end{figure}

Figure \ref{fig:TRAPPIST_e_low_R} shows analogous results for the TRAPPIST-1 e models. The 10\% spot coverage model has water features of similar amplitude (100--150 ppm) to the Archean-Earth--like model, while the latter can have its features muted or even inverted by a 30\% faculae coverage on the star. Once again, the 1.1 $\mu$m water band differs between planetary atmosphere and stellar contamination. We note that the Archean-like model displays methane absorption (about 125 ppm amplitude) around 1.65--1.7 and 2.2--2.3 $\mu$m, which is completely absent from the contamination models. Meanwhile, the normally muted features of the aqua planet model are greatly amplified by 10\% spots on the star, making it appear as though its atmosphere has even stronger absorption than the Archean-like model. Once again, significant radius offsets (about 200 ppm in transit depth this time) are introduced by contamination.

Adding arbitrary amounts of unocculted spots on the star (50\% coverage or more) can lead to very large signals with amplitudes of over 500 ppm. 
\edit1{Such coverage fractions may be possible, at least in terms of coverage by strong magnetic activity, for very active M dwarfs \citep{Saar1985,Reiners2009}.}

We note that obtaining sufficient S/N to detect features in individual planets in a system such as TRAPPIST-1 may prove challenging (e.g. if we look at existing transit observations of the TRAPPIST-1 system such as those of \citet{deWit2016} and \citet{Zhang2018}). This should be less of an issue in the coming era of JWST, as existing studies have shown \citep[e.g.][]{Wunderlich2019}.

As an aside, it is important to mention that we only consider secondary planetary atmospheres with high molecular weight, for which the spectroscopic features are considerably weaker than in the case of atmospheres dominated by $\text{H}_2$. For a clear, low molecular weight atmosphere in a system such as TRAPPIST-1, contamination by stellar activity should not be a cause for concerns.

\subsubsection{High Resolution}

At the spectral resolution of current infrared spectrographs, a planet like TRAPPIST-1 b would have a sufficient RV variation across its transit to obtain a reasonably convincing detection, provided sufficient S/N and atmospheric signal, of course. On the other hand, planet e, while not as challenging as K2-18 b in the RV department, shows a much weaker variation across transit \edit1{than TRAPPIST-1 b}, and thus convincing detections may prove challenging. We therefore test models with $R=200,000$ in this case as well.

Since we model a host star of the same apparent magnitude as TRAPPIST-1, in order to achieve sufficient S/N to look for planetary signals after adding photon noise, we need to simulate visits from an ELT. All attempts at computing RME/CLV corrected correlation maps with photon noise as for observations from a 3.6 m telescope resulted in maps dominated by noise, for all three types of models (stellar activity only, planetary atmosphere only, and both combined).

\begin{figure}[ht]
    \centering
    \includegraphics[width=0.6\textwidth]{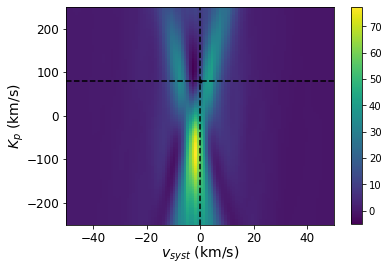}
    \caption{Correlation map (normalized by standard deviation) for the\edit1{ TRAPPIST-1 b} outgassed atmosphere model and 10\% spot coverage on the star with a water line list \citep{Polyansky2018} at 400 K. No correction for RME/CLV was applied. Photon noise was added for 10 ELT visits. \authorcomment1{New figure with the correct $v_{syst}$ and $K_p$ values.}}
    \label{fig:TRAPPIST_B_corr_uncorrected}
\end{figure}

For a cool star like TRAPPIST-1, given the presence of water lines in the stellar spectrum itself, making a convincing water detection requires adequate correction for the RME and CLV. The impact of these effects on correlation maps can clearly be seen in Figure \ref{fig:TRAPPIST_B_corr_uncorrected}: the main positive peak centered around $K_p=-100$ $\text{km s}^{-1}$ and near $v_{\text{syst}}=0$ $\text{km s}^{-1}$ is most likely due to the modulation of stellar line profiles during transit. Our correction method (described in section \ref{section:high_R_analysis}) is thus applied to all the model observations in this analysis.

\begin{figure}[ht]
    \centering
    \includegraphics[width=0.55\textwidth]{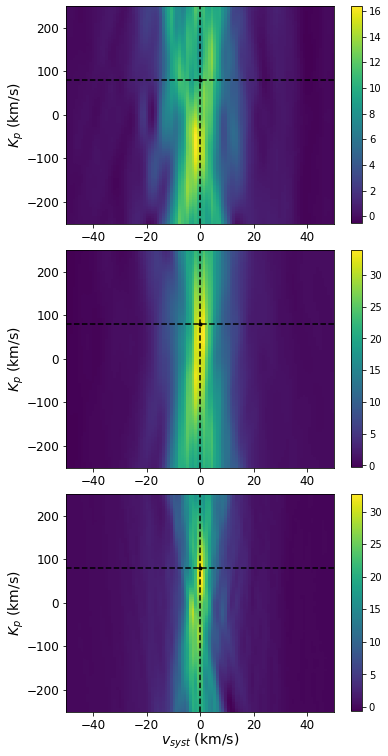}
    \caption{Correlation maps (normalized by standard deviation) for two different TRAPPIST-1 b model transits with a water line list \citep{Polyansky2018} at 400 K. \textit{Top}: Transit with 10\% spot coverage only (no atmosphere). \textit{Middle:} Transit with the outgassed atmosphere model and 10\% spot coverage on the star. \textit{Bottom:} Same as the middle panel but using models at $R=200,000$. All correlation maps were corrected for the RME/CLV signal. In all cases, photon noise was added for 10 ELT visits.}
    \label{fig:TRAPPIST_b_corr}
\end{figure}

In the TRAPPIST-1 b model case, we show in Figure \ref{fig:TRAPPIST_b_corr} that the water signal from the outgassed $\text{O}_2$ atmosphere seems to be detectable with an ELT, provided a sufficient number of visits. In this case, we show that 10 visits appear to be more than enough. In fact, other calculations (not shown in this paper) indicate that as few as five or six visits would suffice to obtain similarly convincing atmospheric detections.
The peak extending between $K_p=0$ and $K_p=-100$ $\text{km s}^{-1}$ in the top panel can only be caused by stellar contamination and residual signal from a possibly incomplete subtraction of the RME/CLV signal. The middle panel shows that, when adding the outgassed atmosphere to the transit, a new peak centered around the true $v_{\text{syst}}$ and planetary $K_p$ appears, in addition to the previous contamination peak. In this case, even though the water contamination signal's strength appears to be of the same order of magnitude as the planetary signal, we can confidently differentiate the origin of both peaks and relate the peak at the planet's $K_p$ to the water in the planet's atmosphere. The correlation with even higher-resolution models ($R=200,000$) in the bottom panel shows a much better distinction between the different signals. We recall from Figure \ref{fig:TRAPPIST_b_low_R} that the water features from the 10 \% spots contamination model used here were degenerate with those of the outgassed model at low resolution. These results show that this degeneracy can be broken with high-S/N, high-resolution observations.

\begin{figure}[ht]
    \centering
    \includegraphics[width=0.55\textwidth]{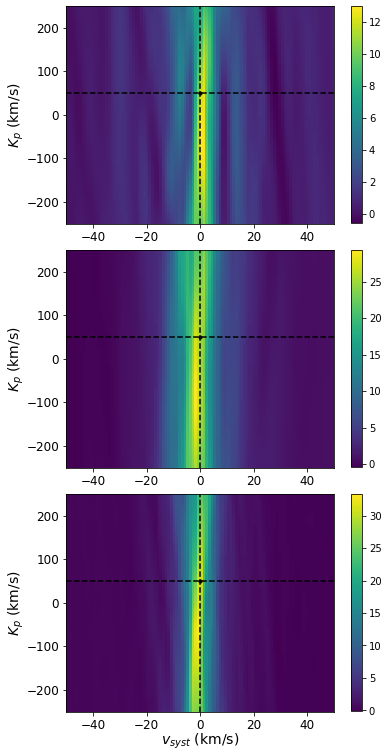}
    \caption{Correlation maps (normalized by standard deviation) for two different TRAPPIST-1 e model transits with a water line list \citep{Polyansky2018} at 300 K. \textit{Top:} Transit with the Archean-Earth--like atmosphere only. \textit{Middle:} Transit with the Archean atmosphere and 10\% spot coverage on the star. \textit{Bottom:} Same as middle but for models at $R=200,000$. All maps were corrected for the RME/CLV signal and photon noise was added as for 30 ELT visits.}
    \label{fig:TRAPPSIT_e_corr}
\end{figure}

The Venus-like model does not produce water features to be detected at high resolution, as was the case at low resolution, so we do not present a correlation map for this atmosphere. This is to be expected considering the low water content and high mean molecular weight of the atmosphere model. However, when adding a 10\% spot coverage to the Venus-like model transits, the contamination signal is effectively detected and the correlation map becomes extremely similar to the map for the pure 10\% spots contamination signal (top panel of Figure \ref{fig:TRAPPIST_b_corr}). Therefore, in this case, high resolution can be used to identify stellar contamination and exclude a false-positive detection made at low resolution.

\begin{figure}[ht]
    \centering
    \includegraphics[width=0.6\textwidth]{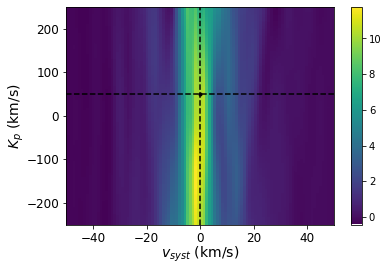}
    \caption{Correlation map (normalized by standard deviation) for a TRAPPIST-1 e model transit with only 10\% spot coverage on the host star (no planetary atmosphere) with a water line list \citep{Polyansky2018} at 300 K. Corrected for RME/CLV. Photon noise was added as for 30 ELT visits.}
    \label{fig:TRAPPIST_e_corr_spots}
\end{figure}

As for the TRAPPIST-1 e case, the Archean-Earth--like atmosphere produces a peak that is difficult to draw conclusions from, even with 30 ELT transits, high S/N values, and even higher spectral resolution ($R=200,000$). The correlation peaks are very extended, especially in the spotted case, and extend to much lower $K_p$ than the true planet velocity, as seen in Figure \ref{fig:TRAPPSIT_e_corr}. The low planet orbital velocity combined with the strength of the contamination signal are likely to blame. As a comparison, Figure \ref{fig:TRAPPIST_e_corr_spots} shows the correlation map for a pure contamination transit. There is a correlation signal present, but it behaves differently from the peaks in Figure \ref{fig:TRAPPSIT_e_corr}: it is broader and slightly off in $v_{\text{syst}}$, and generally stronger at negative values of $K_p$, and may be partially due to an incomplete or inaccurate correction of the RME/CLV contribution.
Finally, the aqua planet model is left out of this analysis because planetary lines are too weak for satisfying correlation peaks to be recovered with noisy models.

\begin{figure}[ht]
    \centering
    \includegraphics[width=0.55\textwidth]{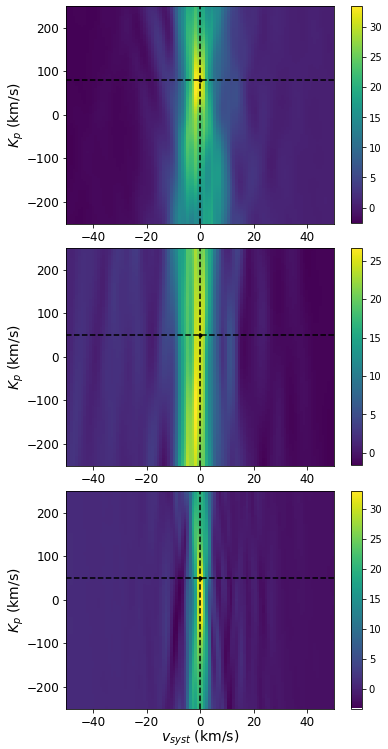}
    \caption{Correlation maps (normalized by standard deviation) for TRAPPIST-1 models with a water line list \citep{Polyansky2018}, in the 1.1--1.2 $\mu$m band. \textit{Top:} Outgassed atmosphere model for TRAPPIST-1 b, with 10\% spot coverage on the star. Noise added for 10 ELT visits. \textit{Middle:} Archean model for TRAPPIST-1 e, with 10\% spot coverage. Noise added for 30 ELT visits. \textit{Bottom:} Same as middle panel, but with spectra at $R=200,000$. RME/CLV correction was applied to all maps. }
    \label{fig:TRAPPIST_corr_maps_new_band}
\end{figure}

Something that can be attempted in order to improve the distinction between planet and stellar contamination signals in this case is to only perform the correlation in an absorption band where stellar contamination is minimal. We have observed for the low-resolution \edit1{TRAPPIST-1} and K2-18 transits that the water band at 1.1--1.2 $\mu$m is not very affected by stellar activity, due to the band being weak at stellar temperatures. With that in mind, we show in Figure~\ref{fig:TRAPPIST_corr_maps_new_band} correlation maps for the TRAPPIST-1 b outgassed and TRAPPIST-1 e Archean atmospheres, both with 10\% spots on the star, correlated with the usual water line list at the planetary temperatures, but this time, in the 1.1-1.2 $\mu$m band. We observe noticeable improvements in the definition of the peaks compared to the equivalent maps in Figures~\ref{fig:TRAPPIST_b_corr} and \ref{fig:TRAPPSIT_e_corr}, indicating that we have indeed filtered out at least part of the stellar contamination signal.

\section{Discussion}
\label{sec:Discussion}

Across the high-resolution analysis, we did not present any results from faculae. Faculae produce inverted features (emission instead of absorption) and therefore a negative correlation signal. It should in principle be easier to single them out if they are a strong contributor to the signal, and they would certainly not be confused with absorption from a planetary atmosphere.

We must point out that the correlation results presented here are definitely best-case scenarios, since we chose to ignore telluric contamination and other sources of noise from the analysis. The detections presented are possible in theory, with the hypothesis that tellurics can be corrected well. The detections presented should also be possible using weaker lines less affected by tellurics, provided additional transits to achieve better S/N.

Based on the difficulties encountered with the models where the planet had small radial velocity variations, we can confidently say that not all targets will be suitable for high-resolution follow-ups, at least with current facilities. In particular, considering that habitable zone planets such as K2-18 b and TRAPPIST-1 e generally have longer orbits than the typical transiting planet, the feasibility of high-resolution transmission spectroscopy of habitable zone planets may prove difficult. Extremely high spectral resolutions (e.g., $R \sim 200,000$) on future spectrographs may alleviate this issue in many cases.

As was also pointed out in other works \cite[e.g.][]{Rackham2018,Rackham2019}, we can notice in the results presented here that, as we go down in spectral type, stellar contamination becomes a larger issue, especially when looking at water lines and features. For earlier spectral types, the effect of contamination is significant for the continuum of transit spectra, i.e. radius offsets and slopes. 
\edit1{In the optical, Na, K, H$\alpha$ and TiO features can be introduced by stellar contamination, and other molecules such as MgH may also be affected.}
This makes high-resolution follow-ups especially relevant for systems with late-type host stars\edit1{, but also interesting for those with early-type stars}.

There is a possibility (which has yet to be verified) that current data reduction methods for transmission spectroscopy already remove some of the contamination signal along with the stellar signal. This may be the case for PCA-based methods \citep{Damiano2019} used to reconstruct the stationary stellar and telluric spectra, for example. In this analysis, we did not use such methods, since the out-of-transit stellar spectrum is already modeled and easy to remove from transit observations, leaving only contamination and RME/CLV signals. 
Since, in principle, activity contamination signals are also stationary in RV space during transit, it may be feasible for PCA algorithms to reconstruct them as well. Whether this is true or not, this sort of possibility is not applicable to low-resolution observations. This is an additional way the use of high-resolution spectroscopy could potentially help correct for the effect of stellar contamination.

This does not mean that there is no need to perform adequate diagnostics of stellar activity such as activity indices and photometric variability when making high-resolution observations of a system, especially for later-type stars, for which spots appear to extend correlation peaks toward $K_p=0$ and $K_p<0$. Of course, it cannot hurt to know more about the host star. Such diagnostics could also help subtracting signals from spots if they are not removed by cleaning the stellar spectrum.

Additionally, constraining active region temperatures and filling factors through other methods may also be very beneficial; possible methods include Doppler imaging \citep{Barnes2015, Barnes2016, Barnes2017}, spot crossing events \citep{Espinoza2019}, stellar spectrum fitting \citep{Zhang2018}, and other methods like chromatic stellar RVs and the chromatic index \citep{Baroch2020}.

The offset of planet radius measurements may be the most challenging effect of stellar contamination to correct for by looking strictly at transit depth data. Alternative methods have been suggested in order to obtain independent radius measurements. In particular, accurate measurements of ingress and egress duration, combined with well constrained orbital parameters, could provide an alternate robust planet radius \citep{Morris2018}.

\begin{figure}[t]
    \centering
    \includegraphics[width=0.65\textwidth]{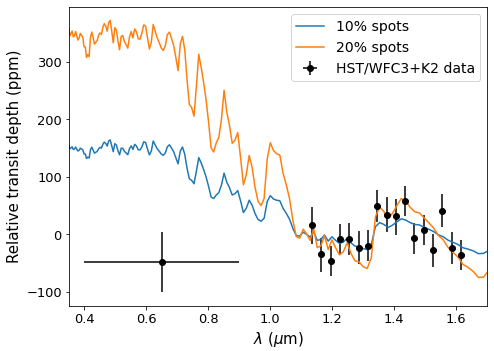}
    \caption{\edit1{Comparison K2-18 type model transits with pure stellar contamination and the existing HST/WFC3 and K2 data sets (black dots), in relative transit depths. The contamination models were taken with cold spots ($T_{\text{spot}}=2600$ K), with 10\% (blue) and 20\% (orange) spot coverage fractions.} \authorcomment1{New figure.} }
    \label{fig:K2-18_contamination_data_comparison}
\end{figure}

We wish to address the recent published detections of atmospheric water vapor in K2-18 b \citep{Benneke2019, Tsiaras2019}. 
\edit1{Using strictly the infrared, it appears that stellar contamination could produce such a detection, or at least strongly influence atmospheric retrievals. However, looking toward the optical, we find that the available K2 photometry \citep{Benneke2019} can offer a discriminant in favor of a true atmospheric detection. Indeed, as seen in Figure~\ref{fig:K2-18_contamination_data_comparison}, contamination models are incompatible with the optical data. Therefore,} 
our \edit1{models} do not invalidate these detections; rather, they support them \edit1{in this specific case}. 
It is still possible, however, that the amplitude of features may have been modulated by stellar activity like in Figure~\ref{fig:K2-18_mixed_models}, which in turn could bias retrievals of atmospheric abundances and structure.

The results in this work provide some justification for ground-based, high-resolution, near infrared spectroscopy follow-ups for many planetary systems, even in the coming era of JWST. This applies especially to systems with late-M dwarf host stars. Stellar contamination affects atmospheric characterization efforts at low and high resolution very differently. At low resolution, abundance retrievals and planet radius measurements are biased, with no obvious or general way to identify the contribution of stellar contamination to the transmission spectrum. In worst-case scenarios, stellar activity could be the only source of significant water absorption detections, or even mute planetary features entirely. Meanwhile, at high resolution, active regions on the star do contribute to the signals detected in correlation, but it is feasible in many cases to use the planet's orbital motion to clearly isolate the planetary atmosphere's signal from the contamination signal. 
In certain cases, it may even be possible to single out bands where the effect of stellar activity is weaker than in other important bands, and further distinguish planetary and stellar signals.
Therefore, combining high-resolution results with existing or future space-based transit results could confirm many detections or non-detections, as well as invalidate tentative or weak low-resolution detections.


\section{Conclusion}
\label{sec:Conclusion}

We have analyzed transit models for three archetype systems at low and high resolution, to examine the effects of stellar activity on transmission spectroscopy and determine if degeneracies at low resolution can be lifted using high-resolution observations. We compared cases with only contamination, cases with only planetary atmospheres, and as mixed cases. 
We found that, at low resolution, transmission spectra can be affected to the point where what appear to be planetary water absorption features are in fact only due to the effect of active regions on the star. This occurs when the molecules responsible for these absorption features are also present in the active regions.
Atmospheric characterization and abundance retrievals can thus suffer strong biases, with no guaranteed general solution using only low-resolution observations. 
However, we find that degeneracies observed at low resolution are in principle breakable at high resolution using cross-correlation maps to separate the planet and stellar contamination correlation peaks. This is true as long as the variation of the planet's radial velocity during transit is sufficient and observations with adequate S/N can be obtained. We observe that, when looking at water lines and features, the effect of stellar contamination grows stronger with later spectral types. We have confirmed the need to correct for the Rossiter--McLaughlin effect and center-to-limb variations when looking for signals from molecules also found in the host star's atmosphere by using CCFs at high resolution.

Numerous improvements to our models should be made in future works. Additional stellar models should be obtained, with multiple goals in mind: to allow for flexible active region temperature contrasts, to model more host star types, and to model late-type stars more accurately using DRIFT-PHOENIX models \citep{Witte2009}. The wavelength domain of the models should be extended to include at least the full JWST spectral domain. Similarly, more realistic, complete, and self-consistent atmosphere models should be used. To better quantify the feasibility of high-resolution cases, tellurics and other sources of noise should be fully included and treated. This would in turn make it easier to apply actual data analysis routines to the models. The geometry of the models must be made more flexible by at least including the spin-orbit misalignment angle $\lambda$ as a model parameter. The rotation of active regions should be implemented during transits as well, in cases where the timescales of transit and stellar rotation are not too different. Cases with occulted active regions need to be fully investigated. For completeness, the reflex RV motion of the star could also be implemented. 

An important goal for the future of this work would be to help in planning efficient observing strategies for potentially problematic targets. With the coming launch of JWST and the current wave of new generation near infrared spectrographs (such as SPIRou, CARMENES, CRIRES+, NIRPS, etc.), all the elements are in place to be able to obtain complementary observations and possibly resolve degeneracies for future targets.

\acknowledgements

This research made use of Photutils, an Astropy package for detection and photometry of astronomical sources \citep{Bradley2019}. This research also used the petitRADTRANS radiative transfer code \citep{Molliere2019} to generate planetary atmosphere models. The Python packages \texttt{emcee} \citep{Foreman-Mackey2013} and \texttt{batman} \citep{Kreidberg2015} were also used. Peter Hauschildt provided specific intensity spectra for stellar models generated with the PHOENIX code \citep{Hauschildt1999}. Molecular line lists were obtained using the HITRAN/HITEMP (\url{https://hitran.iao.ru/molecule/simlaunch}) and ExoMol (\url{http://exomol.com/}) online services. This research was made possible thanks to the funding and support of the Fonds de Recherche du Qu\'ebec - Nature et Technologies (FRQNT), the Natural Sciences and Engineering Research Council of Canada (NSERC), and the Institute for Research on Exoplanets (iREx).

\software{petitRADTRANS \citep{Molliere2019}, Photutils \citep{Bradley2019}, emcee \citep{Foreman-Mackey2013}, batman \citep{Kreidberg2015}, PHOENIX \citep{Hauschildt1999}}


\bibliography{ref}{}
\bibliographystyle{aasjournal}

\end{document}